\documentclass[a4paper, amsfonts, amssymb, amsmath, reprint, showkeys, nofootinbib, twoside]{revtex4-1}
\usepackage[english]{babel}
\usepackage[utf8]{inputenc}
\usepackage[colorinlistoftodos, color=green!40, prependcaption]{todonotes}
\usepackage{subfigure}
\usepackage[pdftex, pdftitle={Article}, pdfauthor={Author}]{hyperref} 
\usepackage[toc]{appendix}
\begin{document}

\title{The TeV gamma-ray luminosity of the Milky-Way and\\
the contribution of H.E.S.S. unresolved sources to VHE diffuse emission}
\author{M. Cataldo$^{1,2}$, G. Pagliaroli$^{2,3}$, V. Vecchiotti$^{2,3}$ and F.L. Villante$^{1,2}$}
\affiliation{$^{1}$University of L'Aquila, Physics and Chemistry Department, 67100 L'Aquila, Italy\\
$^{2}$INFN, Laboratori Nazionali del Gran Sasso, 67100 Assergi (AQ),  Italy\\
$^{3}$Gran Sasso Science Institute, 67100 L'Aquila, Italy}




\begin{abstract}

{H.E.S.S. has recently completed a systematic survey of the Galactic plane in the TeV energy domain.
We analyze the flux, latitude and longitude distributions of $\gamma-$ray sources  
observed by H.E.S.S. in order to infer the properties of Galactic TeV source population.
We show that the total Milky-Way luminosity in the 1-100 TeV energy range is $L_{\rm MW} = 1.7^{+0.5}_{-0.4}\times 10^{37} {\rm ergs}\,{\rm sec}^{-1}$.
%
Evaluating the cumulative flux expected at Earth by the considered population, we show that 
H.E.S.S. unresolved sources provide a relevant contribution to the diffuse Galactic emission. 
Finally, in the hypothesis that the majority of bright sources detected by H.E.S.S. are powered by pulsar activity, like e.g. Pulsar Wind Nebulae or TeV halos, we estimate the main properties of the pulsar population.}
\end{abstract}

\maketitle

\section{Introduction} \label{sec:outline}
The field of TeV astronomy is rapidly evolving thanks to the data obtained by recent experiments.
%
Imaging Atmospheric Cherenkov Telescopes (IACT), like H.E.S.S. \cite{Aharonian:2005kn}, MAGIC \cite{Aleksic:2014lkm} and VERITAS \cite{Weekes:2001pd}, and air shower arrays, such as ARGO-YBJ \cite{Bartoli:2013qxm}, Milagro \cite{Atkins:2004yb} and HAWC \cite{Abeysekara:2015qba}, provided a detailed description of Galactic $\gamma-$ray emission in the energy range $0.1-100\,{\rm TeV}$.
Large scale diffusion emission from different regions of the Galactic plane has been measured by H.E.S.S. \cite{Abramowski:2014vox}, Argo \cite{Argo}, HAWC \cite{HAWC}, and Milagro \cite{Milagro} while catalogues of point-like and extended sources have been recently produced by H.E.S.S. \cite{HGPS} and HAWC \cite{Abeysekara:2017hyn}.
At larger energies ($\sim 100\,{\rm TeV}$ or more), IceCube neutrino telescope has reported the existence of an astrophysical population of neutrinos \cite{Aartsen:2013jdh,Aartsen:2014gkd}.
This signal is believed to be mainly due to extragalactic sources but a subdominant Galactic component, produced by Cosmic Ray (CR) interactions with interstellar gas and/or Galactic TeV sources, should also exist.
The search for this Galactic contribution
is in progress and  potentially within the reach of IceCube experiment \cite{Aartsen:2017ujz,Aartsen:2019fau}.

 Even if the knowledge of our Galaxy in the TeV domain has greatly progressed, several problems remain unsolved.
In most cases,  we are not able to determine whether the observed gamma-ray signals are produced at TeV energies by leptonic or hadronic mechanism.
This limits the possibility to use the gamma-neutrino connection, implied by hadronic production, to estimate the neutrino signal from gamma-ray observed sources.
In addition, we still miss a robust determination of the diffuse $\gamma-$ray flux produced at TeV energies by CR interactions with the gas contained in the Galactic disk.
At these energies, the situation is substantially different than the one observed  at $1-100$ GeV by the Fermi-LAT experiment \cite{TheFermi-LAT:2017pvy,Ackermann:2013fwa} where the CR diffuse emission outshines the contribution of individual sources.
The relatively large diffuse flux measured at TeV by Milagro \cite{Milagro}, HESS \cite{Abramowski:2014vox} and HAWC \cite{HAWC} could be explained either as the cumulative contribution of unresolved sources, see e.g. \cite{Linden} or by considering non conventional CR propagation models characterised by position-dependent transport properties, as e.g. \cite{Pothast:2018bvh}.

Although different astrophysical objects, such as Supernova Remnants (SNRs) and Pulsar Wind Nebulae (PWNe), can generate TeV $\gamma-$rays, we still don't know which (if any) class of sources dominate Galactic emission.
Recent observations of Geminga and PSR B0656+14 by Milagro \cite{Abdo:2009ku} and HAWC \cite{Abeysekara:2017old}, provided evidence for a new class of objects powered by pulsar activity, the so-called TeV halos, that could potentially explain a large fraction of bright TeV sources observed in the Sky \cite{Sudoh:2019lav}.

In this work, we perform a population study of the H.E.S.S. Galactic Plane Survey (HGPS) catalogue with the goal of addressing some of the above open issues.
The HGPS catalogue is particularly useful for our purposes because it provides the optimal sky coverage, encompassing about $\sim 80\%$ of the Galactic plane within its observation region.
We analyze the flux, latitude and longitude distributions of sources detected by H.E.S.S. in order to infer the properties of TeV source population. 
To avoid selection effects, we include in our analysis the brightest sources with a flux above 1 TeV larger than $10\%$ of the CRAB flux. 
By performing a general analysis based on suitable assumptions for the source space and luminosity distributions, we show that the HGPS data permit to estimate with relatively good accuracy the total Milky Way luminosity produced by TeV sources and the total Galactic flux due to both resolved and unresolved sources in the H.E.S.S. Field of View (FoV). The FoV covers in longitude the range $-110^\circ \le l \le 60^\circ$ and in latitudes $|b|<3^\circ$.
This allows us to quantify the contribution of unresolved sources to the total flux, showing that unresolved contribution is possibly the dominant component of the large-scale diffuse signal observed at TeV by H.E.S.S. \cite{Abramowski:2014vox} and Milagro \cite{Atkins:2005wu}.
We then consider the regime where all bright sources observed by H.E.S.S. (which are not firmly identified as SNRs) are powered by pulsar activity, e.g. PWNe and/or TeV halos as suggested by \cite{HGPS} and
we discuss the constraints on the pulsar properties, namely the initial spin period and magnetic field, that are obtained by HGPS data. 
Our analysis of the TeV source population improves and complements previous discussions on the subject, like e.g. that provided by \cite{CasanovaDingus:2007}, by considering different aspects and an original approach and by taking advantage of more recent observational data.

The plan of the paper is as follows. In Sec. \ref{sec:develop} we discuss the HGPS catalogue. In Sec.\ref{sec:method} we present our method to describe the TeV source population. In Sec. \ref{sec:results} we show our results and we discuss their robustness. In Sec. \ref{sec:conclusions} we draw our conclusions.     

\section{H.E.S.S. HGPS} \label{sec:develop}
The H.E.S.S. Galactic Plane Survey (HGPS) catalogue \cite{HGPS} includes 78 VHE sources observed in the longitude range $-110^\circ \le l \le 60^\circ$ and for latitudes $|b|<3^\circ$, measured with an angular resolution of $0.08^\circ$ and a sensitivity $\simeq 1.5\%$ Crab flux for point-like objects.
The integral flux above 1 TeV of each source is obtained from the morphology fit of flux maps, assuming a power-law spectrum with index $\beta=2.3$. 
In order to be consistent with this procedure, we adopt the same assumption to describe the spectrum of galactic sources in the TeV domain. The value $\beta=2.3$ is compatible with the average spectral index obtained by fitting HGPS sources by using a power law or a power-law with an exponential cutoff in the energy range $0.2 \; {\rm TeV} \le E_{\gamma} \le 100\; {\rm TeV}$.

In the following, we focus on the bright sources that produce a photon flux above 1 TeV larger than 10\% of that produced by the CRAB nebula. Above this threshold, the HGPS catalogue can be considered complete \cite{HGPS} and consists of 32 sources: 19 are unidentified, 3 are firmly associated with SNRs (Vela Junior, RCW 86, RX J1713.7-3946), 2 are objects showing evidence of both shell and nebular emission which we refer to as composite objects, and 8 are associated with PWN. 

The HGPS survey provides optimal sky  coverage to perform galactic population studies. Indeed, the observation window $-110^\circ \le l \le 60^\circ$ and $|b|<3^\circ$ includes about $80\%$ of potential sources located in the galactic plane, according to PWNe and SNR distributions parameterized by \cite{Lorimer} and \cite{Green:2015isa}, respectively.
The HAWC experiment reports the longitudinal gamma-ray profile in the angular region $0^\circ < l < 180^\circ$ and $|b|<2^\circ$, for a photon median energy $E_{\gamma} = 7\ {\rm TeV}$ \cite{HAWC}. The Argo-YBJ experiment measures the total gamma-ray emission in the longitudinal region $40^\circ < l < 100^\circ$ and latitudes $|b|<5^\circ$ for $E_{\gamma} = 600\ {\rm GeV}$ \cite{Argo}. At higher energy, $E_{\gamma} = 15\ {\rm TeV}$, the Milagro experiment reports the total gamma-ray emission for longitudes $30^\circ < l < 110^\circ$ and $136^\circ < l < 216^\circ$ and for latitudes $|b|< 10^\circ $ \cite{Milagro}. The sky regions probed by Milagro, Argo-YBJ, and HAWC contain a smaller fraction of the potential sources in the Galactic plane, equal to $\simeq 20 \%$, $\simeq 20\%$, and $\simeq 40\%$, respectively.

\section{Method} 
\label{sec:method}

In order to predict the signal observed by H.E.S.S., we need to consider the space and intrinsic luminosity distribution of the TeV sources. We assume that this can be factorized as the product:
\begin{equation}
\frac{dN}{d^3 r\,dL} = \rho\left({\bf r} \right) Y\left(L\right)  
\label{SpaceLumDist}
\end{equation}
where ${\bf r}$ indicates the position in the Galaxy and $L$ is the $\gamma-$ray luminosity integrated in the energy range $1-100\,{\rm TeV}$ probed by H.E.S.S.. 
The function $\rho({\bf r})$, which is conventionally normalized to one when integrated in the entire Galaxy, is assumed to be proportional to the pulsar distribution in the Galactic plane parametrized by \cite{Lorimer}. The source density along the direction perpendicular to the Galactic plane is assumed to scale as $\exp \left(-\left|z \right|/H\right)$ where $H=0.2\,{\rm kpc}$ represents the thickness of the Galactic disk.

We assume that the intrinsic luminosity distribution $Y(L)$ can be parameterized as a power-law:
\begin{equation}
Y(L)=\frac{{\mathcal N}}{L_{\rm max}}\left(\frac{L}{L_{\rm max}}\right)^{-\alpha}
\label{LumDist1} 
\end{equation}
that extends in the luminosity range $L_{\rm min} \le L\le L_{\rm max}$ \cite{Strong:2006hf}. We take $\alpha = 1.5$ as working hypothesis, since this value can be motivated in the context of sources connected with pulsar activity, such as Pulsar Wind Nebulae (PWNe) and/or TeV halos. 
Other options for the power-law index $\alpha$ (and other assumptions in the analysis) will be also considered, see Tab.\ref{tabres}, in order to test the stability of our results.

The parameter ${\mathcal N}$ defined in Eq.(\ref{LumDist1}) determines the high-luminosity normalization of the function $Y(L)$; it represents the number of sources per logarithmic luminosity interval at the maximal luminosity (i.e. $dN/d\ln L = {\mathcal N}$ for $L=L_{\rm max}$); its physical meaning in the context of a fading source population is discussed in the next section.

The last necessary ingredient to predict the expected signal in H.E.S.S. is the relationship between the intrinsic luminosity $L$ of sources and the flux produced at Earth, that can be generally written as:  
\begin{equation} 
    \Phi=\frac{L}{4 \pi r^2 \langle E \rangle}
\label{Phi}
\end{equation}
where $r$ is the source distance and $\langle E \rangle$ is the average energy of photons emitted in the range $1-100\,{\rm TeV}$. In our calculations, we take the average spectrum observed by HESS as a reference \cite{HGPS}, i.e. we assume that all sources can be described by a power-law in energy with spectral index $\beta = - 2.3$ that corresponds to $\langle E \rangle = 3.25 \,{\rm TeV}$.  

In our analysis, we determine the maximal luminosity $L_{\rm max}$ and the normalization ${\mathcal N}$ of the luminosity function by fitting H.E.S.S. observational results.
This approach is original and different from previous studies on the subject  \cite{CasanovaDingus:2007} where the value of the maximal luminosity is instead assumed "a priori".
The determination of $L_{\rm max}$ and ${\mathcal N}$ allow us to estimate the total TeV luminosity produced by the considered population in the entire Galaxy which is given by:
\begin{equation}
L^{\rm MW}= 
\frac{{\mathcal N}L_{\rm max}}{\left(2-\alpha\right)} 
\left[1 -\Delta^{\alpha-2}\right]
\label{Lmw}\end{equation}
where $\Delta \equiv L_{\rm max}/L_{\rm min}$.
The minimal luminosity $L_{\rm min}$ cannot be constrained by HESS observations. However, its value marginally affects the quantities considered in this paper, provided that $\Delta \gg 1$. 
Unless otherwise specified, we quote the results obtained for $\Delta \to \infty$ that can be easily recalculated by using the above equation, if other values are considered.

By using Eqs.(\ref{SpaceLumDist},\ref{LumDist1},\ref{Phi}), we can also calculate the flux at Earth produced by all sources (resolved and not resolved) included in the H.E.S.S. Field of View (FoV). This can be expressed as:
\begin{equation}
\Phi_{\rm tot} = 
\xi\;
\frac{L_{\rm MW}}{4\pi\langle E \rangle}\; 
\langle r^{-2} \rangle
\label{phitot}
\end{equation}
where the parameter $\xi$, which is defined as
\begin{equation}
\xi \equiv \int _{\rm FoV}d^3r \, \rho({\bf r}) = 0.812 ,
\end{equation}
represents the fraction of sources of the considered population which are included in the H.E.S.S. FoV while the quantity $\langle r^{-2} \rangle$, defined as:
\begin{equation}
\langle r^{-2} \rangle \equiv \frac{1}{\xi}
\int_{\rm FoV}d^3r \, \rho({\bf r}) \; r^{-2} = 0.0176 \,{\rm kpc}^{-2}
\end{equation}
is the average value of their inverse square distance. 
While the above values are specific for HGPS survey (and for the adopted source spatial distribution $\rho(\bf r)$), Eq.(\ref{phitot}) has a general validity; it can be used to evaluate the expected flux in a generic experiment and for an arbitrary source distribution, provided that the corresponding $\xi$ and $\langle r^{-2} \rangle$ are coherently calculated.

\subsection{Pulsar Wind Nebulae}
\label{subsec:PWN}

 The luminosity distribution given in Eq.(\ref{LumDist1}) can be naturally obtained by assuming a population of {\em fading} sources with intrinsic luminosity that decreases over a time scale $\tau$ according to:
\begin{equation}
L(t)= L_{\rm max}\left(1+\frac{t}{\tau}\right)^{-\gamma}
\label{lum}
\end{equation}
where $t\le T_{\rm d}$ indicates the time passed since source formation, $T_{\rm d}$ is the total duration of TeV-emission and $L_{\rm max}$ is the initial luminosity.
If we assume that the birth-rate $R$ of these sources in the Galaxy is constant in time, we can calculate the luminosity function $Y(L)$ that is given by:
\begin{equation}
Y(L) = \frac{R \, \tau \, (\alpha - 1)} {L_{\rm max}}\left(\frac{L}{L_{\rm max}}\right)^{-\alpha}
\label{LumDist}
\end{equation}
where $\alpha = 1/\gamma + 1$ and $L_{\rm min} \equiv L(T_{\rm d})$. 
In this assumption, the normalization factor ${\mathcal N} = R \, \tau \, (\alpha-1)$ of the luminosity distribution has a precise physical meaning; it basically represents the total number of {\em young} sources in the Galaxy that had not enough time to loose their initial luminosity and that are expected to be more easily detected by H.E.S.S.. Note that, the observational determination of ${\mathcal N}$ can be converted into a bound on the fading timescale $\tau$, if the source formation rate is known.

The above description can be applied to potential TeV sources in the Galaxy, such as PWNe \cite{Gaensler:2006ua} or TeV Halos \cite{Linden}, which are connected  with the explosion of core-collapse SN and the formation of a pulsar. 
The birth rate of these objects can be assumed proportional to that of SN explosions in our Galaxy, i.e. $R_{\rm SN} = 0.019\,{\rm yr}^{-1}$ as recently measured by \cite{Diehl:2006cf}. We thus write $R=\varepsilon \, R_{\rm SN}$ assuming $\varepsilon = 1$ for simplicity, unless otherwise specified. 
If the TeV-emission is powered by pulsar activity it is reasonable to assume that TeV-luminosity is proportional to the pulsar spindown power, i.e. 
\begin{equation}
L = \lambda\, \Dot{E}
\end{equation}
where $\lambda \le 1$ and:
\begin{equation}
\Dot{E} = \Dot{E}_0 \left(1+\frac{t}{\tau_{\rm sd}}\right)^{-2}
\end{equation}
with:
\begin{eqnarray}
\nonumber
\Dot{E}_0 &=&\frac{8\pi^4 B_0^2 R^6}{3 c^3 P_0^4}\\
\tau_{\rm sd} &=& \frac{3 I c^3 P_0^2}{4\pi^2B_0^2 R^6}
\end{eqnarray}
where $P_0$ and $B_0$ are the initial spin period and magnetic field \cite{Shapiro} while 
the inertial momentum is $I = 1.4\cdot 10^{45} \,{\rm g\, cm}^{2}$ and the pulsar radius $R= 12 \,{\rm km}$ \cite{Lattimer:2006xb}. 
This implies that the fading timescale is determined by the pulsar spindown time scale, i.e.  $\tau = \tau_{\rm sd}$. 
Moreover, if the efficiency of TeV emission does not depend on time ($\lambda\sim {\rm const}$), the exponent in Eq.~\eqref{lum} is $\gamma=2$, motivating our working hypothesis that the luminosity distribution scales as $Y(L)\propto L^{-1.5}$. 
%
Finally, $P_0$ and $B_0$ can be determined from $L_{\rm max}$ and $\tau$ by using:
\begin{eqnarray}
\nonumber
\frac{P_0}{\rm 1\, ms} &=& 94 
\left(\frac{\lambda}{10^{-3}}\right)^{1/2}  
\left(\frac{\tau}{10^{4}{\rm y}}\right)^{- 1/2}
\left(\frac{L_{\rm max}}{10^{34}{\rm erg\, s}^{-1}}\right)^{- 1/2}  \\
\nonumber
\frac{B_0}{10^{12}{\rm G}} &=&  5.2
\left(\frac{\lambda}{10^{-3}}\right)^{1/2}  
\left(\frac{\tau}{10^{4}{\rm y}}\right)^{- 1}
\left(\frac{L_{\rm max}}{10^{34}{\rm erg\, s}^{-1}}\right)^{- 1/2}\\
\label{P0B0}
\end{eqnarray}
provided that the fraction $\lambda$ of the spin-down power that is converted into TeV $\gamma-$ray emission is known. 

The parameter $\lambda$ is highly uncertain; it is determined by the conversion of the spin-down energy into $e^{\pm}$ pairs (that can be very efficient, see e.g. \cite{Sudoh:2019lav,Manconi:2020ipm}) and by the subsequent production of TeV photons. The values obtained for firmly identified PWNe in the HPGS catalogue are included between $5\times10^{-5}$ and $6\times 10^{-2}$, see Tab.~1 of \cite{Abdalla:2017vci}.
For comparison, the value $\lambda\sim 3\times 10^{-3}$ is obtained in \cite{Linden} by studying the TeV $\gamma-$ray emission of Geminga. 
In this work, we consider $\lambda$ as a free parameter, taking the value $\lambda=10^{-3}$ as a reference in numerical calculations.


The possibility of $\lambda$ being correlated to the spindown power, i.e. $\lambda = 
\lambda_0 ({\dot E}/{\dot E_0})^{\delta}$, is suggested by the results of \cite{Abdalla:2017vci} that found $L = \lambda\,{\dot E} \propto \dot{E}^{1+\delta}$ with $1 + \delta = 0.59 \pm 0.21$ by studying a sample of PWNe in the HPGS catalogue. In this case, one obtains $\gamma \simeq 1.2$ in Eq.~\eqref{lum} that corresponds to a source luminosity function $Y(L)\propto L^{-1.8}$.  
This scenario is also discussed in our analysis and does not introduce relevant changes in our conclusions.
%
The initial spin period $P_0$ and magnetic field $B_0$ can still be derived from Eqs.(\ref{P0B0}) by using the value $\lambda_0$ referred to initial efficiency of TeV emission.

Finally, we consider the effects of dispersion of the initial period and magnetic field around reference values indicated as ${\widetilde P}_0$ and ${\widetilde B}_0$. This in turn implies a dispersion in $L_{\rm max}$ and $\tau$.
The source luminosity function can be obtained by integrating Eq.(\ref{LumDist}), calculated by assuming $\tau=\tau_{\rm sd}(B_0,P_0)$ and $L_{\rm max} = \lambda \,{\dot E}_0(B_0,P_0)$, over $B_0$ and $P_0$ probability distributions. We obtain:
\begin{equation}
Y(L) = \frac{R \, \widetilde{\tau} \, (\alpha - 1)} {\widetilde{L}}\;
\left(
\frac{L}{\widetilde{L}}\right)^{-\alpha}
G\left(\frac{L}{\widetilde{L}}\right)
\label{LumDist2}
\end{equation}
where $\widetilde{\tau}\equiv \tau_{\rm sd}(\widetilde{B}_0,\widetilde{P}_0)$ and 
$\widetilde{L}\equiv L_{\rm max}(\widetilde{B}_0,\widetilde{P}_0)$
are the spin-down timescale and maximal luminosity for the reference values ${\widetilde P}_0$ and ${\widetilde B}_0$.
The obtained luminosity function differs from Eq.(\ref{LumDist}) for the presence of the function $G(L/\widetilde{L})$ that is defined according to:
\begin{equation}
G(x) \equiv 
\int dp \; h(p) p^{6-4\alpha} 
\int db \; g(b) b^{2\alpha-4}\,
\theta\left(p^{-4}\, b^{2}-x\right)
\end{equation}
where $p\equiv P_0/{\widetilde P}_0$, $b\equiv B_0/{\widetilde B}_0$, while $h(p)$ and $g(b)$ describe the probability distributions of initial period and magnetic field. 
We assume that these functions 
can be modelled as gaussian distributions in $\log_{10}(p)$ and $\log_{10}(b)$, centered in zero and having widths given by $\sigma_{\log{P}} = \log_{10}(f_p)$ and $\sigma_{\log{B}} = \log_{10}(f_b)$ with the parameters $f_{p}$ and $f_{b}$ described in next section. 
Under this assumption, the parameters $\widetilde{\tau}$ and $\widetilde{L}$ represent the central values of the log-normal (correlated) distributions of $\tau$ and $L_{\rm max}$ that are obtained as a result of the introduction of $P_0$ and $B_0$ dispersions.

\section{Results} 
\label{sec:results}

Flux, latitude, and longitude distributions of sources observed in HGPS are fitted by using an unbinned likelihood (see Appendix \ref{sec:appendix} for details) with the goal of constraining the source luminosity distribution.
In order to avoid selection effects, we restrict our analysis
to the brightest sources that produce a photon flux above 1 TeV larger than 10\% of that produced by the CRAB nebula $\phi_{\rm CRAB}=2.26\times10^{-11}\,{\rm cm}^{-2}\,{\rm s}^{-1}$. Above this threshold, the catalogue consists of 32 sources (3 of which identified as SNRs) and is considered complete \cite{HGPS}.
This allows us to perform our analysis in full generality without being forced to hypothesize a prescribed physical dimension for the sources because the angular extension does not discriminate the possible identification. 
A possible exception is provided by very close and very extended sources that cover angular regions larger than $\sim 1^\circ$ and could escape detection due to background subtraction procedure employed by H.E.S.S.
We checked, however, that this situation is unlikely and, thus, does not affect our constraints unless one assumes that the majority of the observed sources have physical extension much larger than $few \times 10\,{\rm pc}$.    
In conclusion, the obtained results may be applied to PWNe 
as well as to TeV halos, provided that they have dimension that do not exceed $\sim 40\,{\rm pc}$.

\begin{figure}[t]
$$
\includegraphics[width=0.5\textwidth]{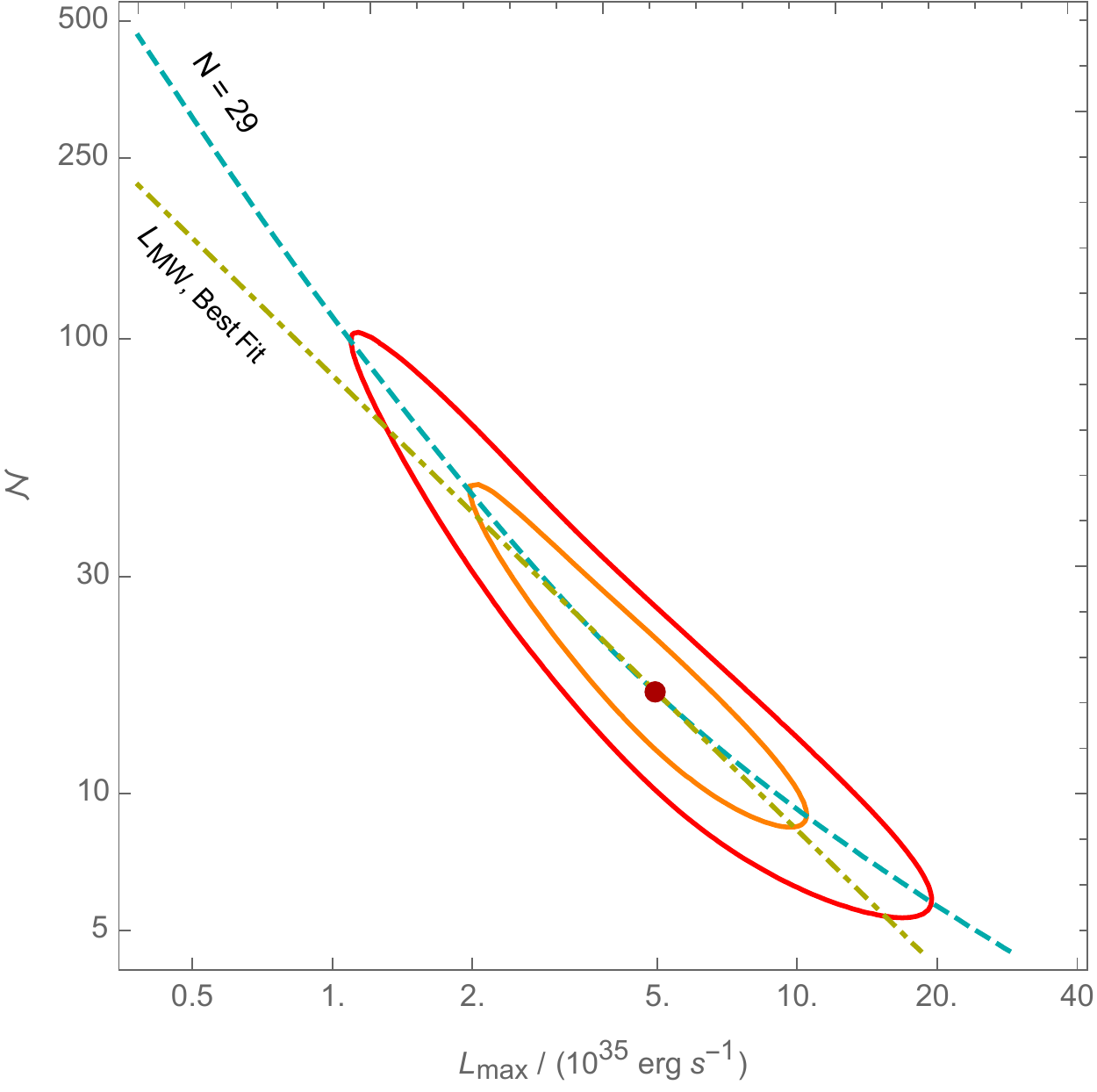}
$$
\vspace{-0.9cm}
\caption{\em 
The best fit and the 1 and 2 $\sigma$ allowed regions for the maximal luminosity $L_{\rm max}$ and the normalization ${\mathcal N}$ of the luminosity distribution of galactic TeV sources. \label{fig1}}
\label{LmaxN}
\end{figure}

The best fit values and the allowed regions for the maximal luminosity $L_{\rm max}$ and the  normalization ${\mathcal N}$ of the source luminosity distribution are shown in Fig.\ref{LmaxN}. We obtain:
\begin{eqnarray}
\nonumber 
L_{\rm max} &=&  4.9^{+3.0}_{-2.1} \times 10^{35} {\rm ergs/s}  \\
{\mathcal N} &=&17^{+14}_{-6}
\label{LmaxNBF}
\end{eqnarray}
where the quoted uncertainties correspond to $1\sigma$ confidence level (CL).
The constraint on the maximal luminosity can be also expressed as $L_{\rm max} =  13^{+8}_{-6} \, L_{\rm CRAB}$ by considering that the CRAB luminosity (above 1 TeV) is $L_{\rm CRAB}=3.8\cdot 10^{34} {\rm ergs/sec}$.
The above results are obtained for our reference case where we assume that the source distribution is proportional to that of pulsars given by \cite{Lorimer}, the disk thickness is $H=0.2\,{\rm kpc}$ and the power-law index of the luminosity distribution is $\alpha=1.5$. 
Moreover, we include 29 HPGS sources neglecting the 3 sources which are firmly identified as SNRs.
This is motivated by the fact that we discuss, in next section, the possible interpretation of our results in terms of a population of fading sources powered by pulsar activity. 
The dependence and/or stability of the obtained results with respect to this and other assumptions in our analysis are discussed in details in Tab. \ref{tabres} and further commented at the end of this section. 

\begin{figure}[t]
$$
\includegraphics[width=0.5\textwidth]{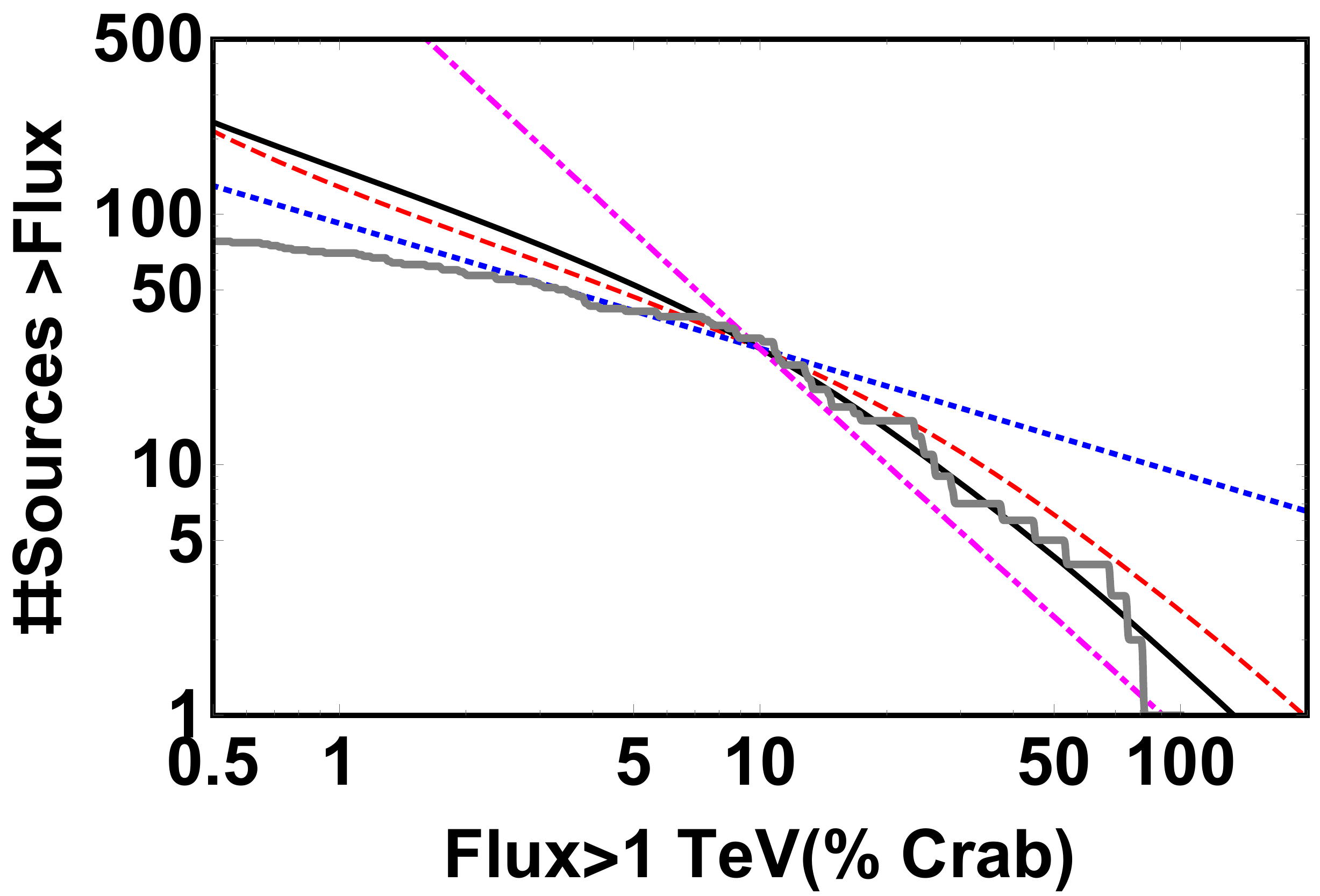}
$$
\vspace{-0.9cm}
\caption{\em The cumulative distribution of the HGPS sources (gray line) compared with expectations for different values of the maximal luminosity $L_{\rm max}$. 
The black line is obtained for the best-fit values in Eq.\eqref{LmaxNBF}, the magenta dot-dashed line and the blue dotted one are obtained for the limit cases of $L_{\rm max}\to 0$ and $L_{\rm max}\to \infty$, respectively, while the red dashed line shows an intermediate case of $L_{\rm max}=30\times L_{\rm CRAB}$.  \label{fig2}}
\end{figure}

The obtained bounds are connected with specific features of the H.E.S.S. data.
The constraint on the maximal luminosity essentially originates from the flux distribution of HGPS sources, as can be understood by looking at Fig.\ref{fig2} where we compare the cumulative number $N(\Phi)$ of observed sources with a flux larger than $\Phi$ with the predictions obtained for different $L_{\rm max}$ values.
The theoretical calculations are normalized in such a way that the expected number of sources with $\Phi\ge 0.1 \Phi_{\rm CRAB}$ is equal to the observational value $N_{\rm obs}=29$. This corresponds to moving along the cyan dashed line in Fig.~\ref{LmaxN} that maximizes the likelihood for each assumed $L_{\rm max}$.
The black line in Fig.~\ref{fig2} corresponds to the best fit value $L_{\rm max}=13\;L_{\rm CRAB}$ and well reproduces the flux distribution in the range $\Phi\ge 0.1\Phi_{\rm CRAB}$ considered in our analysis. 
For comparison, we also show with a red dashed line the expected behaviour of $N(\Phi)$ for $L_{\rm max} = 30\,L_{\rm CRAB}$. This value is disfavoured at $\sim 2\sigma$ level by HGPS data because bright sources are overproduced with respect to observational results.

A more complete understanding of the above points can be obtained by considering the magenta dot-dashed line and the blue dotted line in Fig.~\ref{fig2} that correspond to the limiting cases $L_{\rm max}\to 0$ and $L_{\rm max}\to \infty$, respectively. 
For both these assumptions, the flux distribution can be derived analytically, as it is discussed in the  \ref{sec:appendix}.
Namely, for $L_{\rm max}\to \infty$, the source flux distribution $dN/d\Phi$ is described by a power-law with the same index of the luminosity function, so that the cumulative distribution scales as $N(\Phi)\propto \Phi^{1-\alpha}$.
When $L_{\rm max}\to 0$, one instead obtains  $dN/d\Phi\propto \Phi^{-5/2}$, predicting $N(\Phi) \propto \Phi^{-3/2}$ independently from the assumed source luminosity function.
The cumulative distribution of sources observed by H.E.S.S. has a different behaviour with respect to both cases and thus it requires a specific $L_{\rm max}$ value in order to be reproduced.
The possibility to determine $L_{\rm max}$ from the flux distribution automatically implies the ability to fit the normalization ${\mathcal N}$ of the source luminosity function by considering the additional constraint provided by the total number of observed sources, as it is understood by looking at the cyan dashed line in Fig.\ref{LmaxN}.

By using Eqs.~(\ref{Lmw}) and (\ref{phitot}), we  obtain a determination of the total luminosity of the Galaxy in the energy range $1-100\,{\rm TeV}$ and of the total flux (in the same energy range) produced by sources in the H.E.S.S. FoV. We get:
%
%
%
\begin{eqnarray}
\nonumber
L_{\rm MW} &=& 1.7^{+0.5}_{-0.4}\times 10^{37} {\rm ergs}\,{\rm sec}^{-1}\\
\Phi_{\rm tot} &=& 3.8^{+1.0}_{-1.0}\times 10^{-10} {\rm cm}^{-2}\, {\rm sec}^{-1}
\label{BFLum}
\end{eqnarray}
that correspond to $L_{\rm MW} =445^{+138}_{-112}\, L_{\rm CRAB}$ and 
$\Phi_{\rm tot}= 16.8^{+4.4}_{-3.5}\, \Phi_{\rm CRAB}$ in CRAB units.
We note that the errors on these quantities are relatively small because they are proportional to the product ${\mathcal N}\,L_{\rm max}$ that is well constrained by observational data, as it is also understood by considering the green dot-dashed line in Fig.\ref{LmaxN}.
The total TeV luminosity is only a factor $\sim 4$ smaller than that obtained in the energy range $1-100\,{\rm GeV}$ by fitting the Fermi-LAT 3FGL \cite{TheFermi-LAT:2017pvy} and 1FHL \cite{Ackermann:2013fwa} catalogues. 

The total flux at Earth $\Phi_{\rm tot}$ should be compared with the cumulative emission produced by all the 78 resolved sources in the HGPS catalogue, i.e. $\Phi_{\rm HGPS}= 10.4\, \Phi_{\rm CRAB}$.
We obtain by subtraction the unresolved flux $\Phi_{\rm NR}= 7.7^{+4.4}_{-3.5}\,\Phi_{\rm CRAB}$ which is due to sources in the considered population that are too faint to be identified by H.E.S.S..
We see that unresolved emission $\Phi_{\rm NR}$ is relatively large, comparable to the resolved source contribution. 
This is naturally expected because the observational horizon for H.E.S.S. is limited, while sources are expected to be distributed everywhere in the Galaxy\footnote{ 
As an example, a source with intrinsic luminosity $L\simeq L_{\rm CRAB}$ produces a flux larger than $0.1 \Phi_{\rm CRAB}$, only at a distance smaller than $r \simeq 6\;{\rm kpc}$.}.
In agreement with our previous estimate of this quantity \cite{Cataldo2019}, we obtain $\Phi_{\rm NR}\simeq 60\%\,\Phi_{\rm HGPS}$.   

\begin{table*}[th]
\begin{tabular}{l|ccccccc}
& $\log_{10}\frac{L_{\rm max}}{{\rm erg\, s^{-1}}}$ &${\mathcal N}$& $\log_{10}\frac{L_{\rm MW}}{{\rm erg\ s^{-1}}}$ & $\Phi_{\rm tot}$
& $\tau$
& $\Delta \chi^2$\\
 \hline
 Ref.    & $35.69^{+0.21}_{-0.28}$ & $17^{+14}_{-6}$ & $37.22^{+0.12}_{-0.13}$ & $3.8^{+1.0}_{-1.0}$ & $1.8^{+1.5}_{-0.6}$ & $-$\\
SNR & $35.69^{+0.22}_{-0.25}$ & $18^{+15}_{-7}$ &$37.23^{+0.12}_{-0.13}$ &$3.8^{+1.0}_{-1.0}$ & $1.8^{+1.6}_{-0.7}$ & $1.4$ \\
 $H=0.1\,{\rm kpc}$ & $35.65^{+0.22}_{-0.27}$ & $15^{+14.5}_{-6}$ & $37.13^{+0.12}_{-0.13}$& $5.0^{+0.4}_{-2.0}$ & $1.6^{+1.5}_{-0.6}$ & $-7.3$ \\
$d = 20\ {\rm pc}$ & $35.69_{-0.26}^{+0.20}$ & $17_{-6}^{+16}$ & $37.23^{+0.12}_{-0.13}$ & $3.9^{+0.8}_{-1.0}$ & $1.9^{+1.9}_{-0.7}$ & $-0.2$\\
$d = 40\ {\rm pc}$ & $35.67^{+0.20}_{-0.25}$ & $20^{+20}_{-8}$ & $37.28^{+0.12}_{-0.13}$ & $4.4^{+1.2}_{-1.1}$ & $2.2^{+2.0}_{-0.8}$ & $-1.8$\\
 $\alpha=1.3$    & $35.61^{+0.18}_{-0.27}$ & $25^{+24}_{-8.5}$ & $37.17^{+0.12}_{-0.13}$ & $3.5^{+1.1}_{-0.9}$ & $4.3^{+4.3}_{-1.5}$ & $0.0$ \\
 $\alpha=1.8$    & $35.83^{+0.29}_{-0.24}$ & $7^{+6}_{-4}$ & $37.39^{+0.11}_{-0.13}$ & $5.9^{+1.8}_{-0.1}$ & $0.5^{+0.4}_{-0.2}$  & $0.5$\\
  \hline
$N_{\rm obs} = 32$ & $35.71^{+0.22}_{-0.24}$ & $18^{+14}_{-7}$ &$37.26^{+0.12}_{-0.12}$ & $4.2^{+1.3}_{-1.0}$ & $-$ & $-$\\
\hline
\end{tabular}
\caption{\em 
The best fit values and the $1\sigma$ allowed ranges for the maximal luminosity ($L_{\rm max}$); the normalization factor of the luminosity function (${\mathcal N}$); the total TeV Milky Way luminosity ($L_{\rm max}$); the total flux in the H.E.S.S. FoV ($\Phi_{\rm tot}$, expressed in $10^{-10}\,{\rm cm}^{-2}\,{\rm s}^{-1}$); the fading timescale ($\tau$, expressed in ${\rm ky}$). 
The different cases are described in the text. The $\Delta \chi^2$ is calculated respect to our reference case (first row in the table.)}
\label{tabres}
\end{table*}

In conclusion, our results show that unresolved sources are likely to provide a relevant contribution to the diffuse large-scale $\gamma-$ray signal observed by H.E.S.S. and other experiments, with profound implications for the interpretation of observational results in the TeV domain. 
The unresolved flux $\Phi_{\rm NR}$ is comparable to or larger than expectations for the truly diffuse contribution produced by the interaction of high-energy cosmic rays (CR) with the gas contained in the galactic disk. 
This diffuse component can be estimated as $\Phi_{\rm diff} = (5-15)\,\Phi_{\rm CRAB}$ by following the approach of \cite{Cataldo2019,Pagliaroli:2016lgg}, depending on the assumed CR space and energy distribution. 
The estimate $\Phi_{\rm diff} \simeq 15\,\Phi_{\rm CRAB}$ is e.g. obtained by assuming CR spectral hardening toward the galactic center, as recently emerged from analysis of Fermi-LAT data at lower energies \cite{Acero:2016qlg}. 
It was noted in \cite{Cataldo2019} that, 
if unresolved contribution is large (namely, $\Phi_{\rm NR}\ge 0.5\Phi_{\rm HPGS}$),
this possibility is disfavoured by H.E.S.S. \cite{Abramowski:2014vox} because the total flux (resolved + unresolved + truly diffuse signal) obtained in this hypothesis exceeds the total observed emission from the galactic plane. 
%
%
Here, we strengthen this conclusion by noting that the total flux measured by Milagro at 15 TeV ($d\Phi/dE \sim 2.9\times 10^{-12}\,{\rm cm}^{-2}\,{\rm s}^{-1}\,{\rm sr}^{- 1}\,{\rm TeV}^{-1}$ for $30<l<65$ and $|b|<2$) is consistent (within uncertainties, see next section) with the total flux produced by the HGPS source population in the same observation window ($d\Phi^{\rm M}_{\rm HGPS}/dE \sim 3.4 \times 10^{-12}\,{\rm cm}^{-2}\,{\rm s}^{-1}\,{\rm sr}^{1}\,{\rm TeV}^{-1}$). This
suggests that the anomalous diffuse emission reported by Milagro is due to unresolved sources and provides an additional constraint to the possibility of a large truly-diffuse contribution produced by CR interactions in the galactic disk.

\subsection{Robustness of the results}
\label{subsec:robustness}
In the following we briefly discuss the stability of our results with respect to the assumptions adopted in our analysis.
In Tab.\ref{tabres} we consider different scenarios identified by the ingredient which has been modified with respect to the reference case (e.g. the space distribution, the disk thickness, the source physical dimension, the power-law index of the luminosity distribution, etc.).
For each case, we give the best-fit results and the $1\sigma$ allowed regions for the source luminosity function parameters (${\mathcal N}$ and $L_{\rm max}$), the total TeV luminosity of the Galaxy $L_{\rm MW}$, the total flux produced at Earth $\Phi_{\rm tot}$, the fading time scale $\tau$ and the level of agreement with data, expressed in terms of the $\Delta \chi^2$ with respect to our reference case. 

We see that the inclusion of the three sources firmly identified as SNRs in the HPGS catalogue (case labelled as $N_{\rm obs}$=32 in Tab.\ref{tabres}) does not alter our conclusions, marginally affecting the maximal luminosity $L_{\rm max}$ and increasing by less than $10\%$ the normalization ${\mathcal N}$ of the source luminosity distribution. 
No significant effects are produced by assuming that sources follow the SNR distribution parameterized by \cite{Green:2015isa} (case labelled as SNR) instead of the pulsar distribution of \cite{Lorimer}.
The results of our analysis are also unchanged when we modify the thickness of the Galactic disk. 
However, the quality of the fit is substantially improved ($\Delta \chi^2 \simeq - 7$) if we assume a smaller disk thickness ($H=0.1\,{\rm kpc}$) than our reference choice $0.2\ {\rm kpc}$. 
This is due to the fact that the latitudinal distribution of HPGS sources is quite narrow, having a rms latitude of $0.017$, as it expected for a population of young sources connected with the site of past core-collapse supernova explosions. 
In particular, this information can be used in favor of a fading sources population, as young pulsar wind nebulae, not old enough to drift off the galactic plane \cite{Abdalla:2017vci}. This specific hypothesis and its implications will be further discussed in the next section. 

The cases labelled as {$d=20\,{\rm pc}$} and {$d=40\,{\rm pc}$} are obtained by assuming that all sources in the Galaxy have a prescribed physical dimension and that objects with angular extension larger than $\sim 1^\circ$ are not observed by H.E.S.S. We see that our results are not modified in this assumption. 

Finally, we consider the effects produced by a variation of the power index $\alpha$ of the luminosity distribution by considering two cases: $\alpha = 1.3$, $\alpha = 1.8$. 
We obtain a $\sim 10\%$ decrease ($\sim 50\%$ increase) of the TeV Milky way luminosity and of the total flux at Earth for $\alpha=1.3$ ($\alpha=1.8$), with a slight preference for the case with power index $1.3$.

In conclusion, the cumulative sources contribution to the Milky Way luminosity in the energy range $[1,100]\,{\rm TeV}$ and to the total $\gamma-$ray flux in the H.E.S.S. FoV are included in the ranges: $L_{\rm MW} = \left(1.4 - 2.5 \right) \times 10^{37}\ {\rm erg\ s^{-1}}$, $\Phi_{\rm tot}=\left( 3.5 - 5.9\right) \times 10^{-10}\ {\rm cm^{-2}\ s^{-1}}$,
showing that both these quantities can be constrained within a factor of $1.8$ by present observational data.

\begin{figure*}
     \centering
     \begin{subfigure}
        \centering
        \includegraphics[width=8cm]{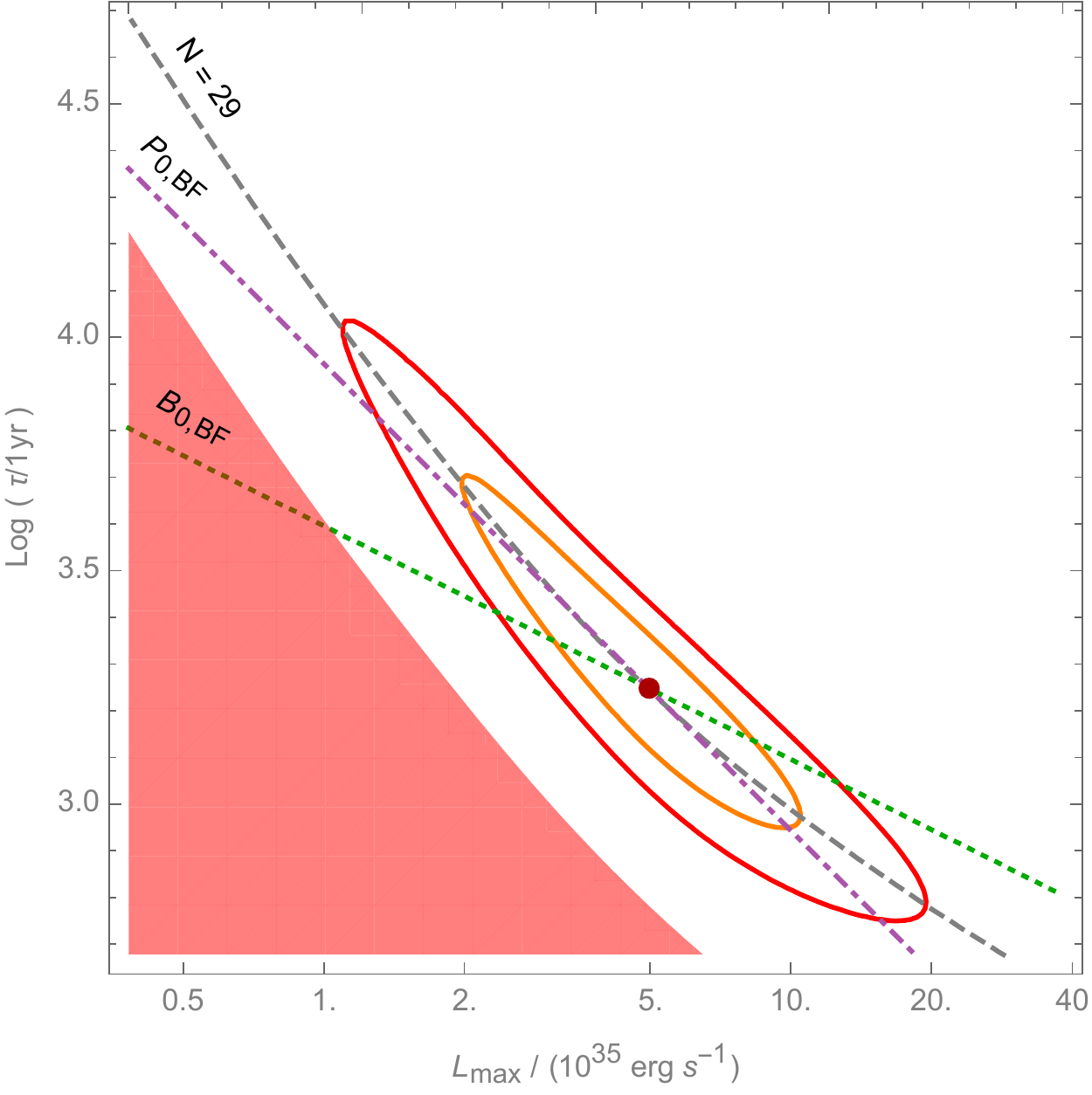}
     \end{subfigure}
     \begin{subfigure}
         \centering
         \includegraphics[width=8.3cm]{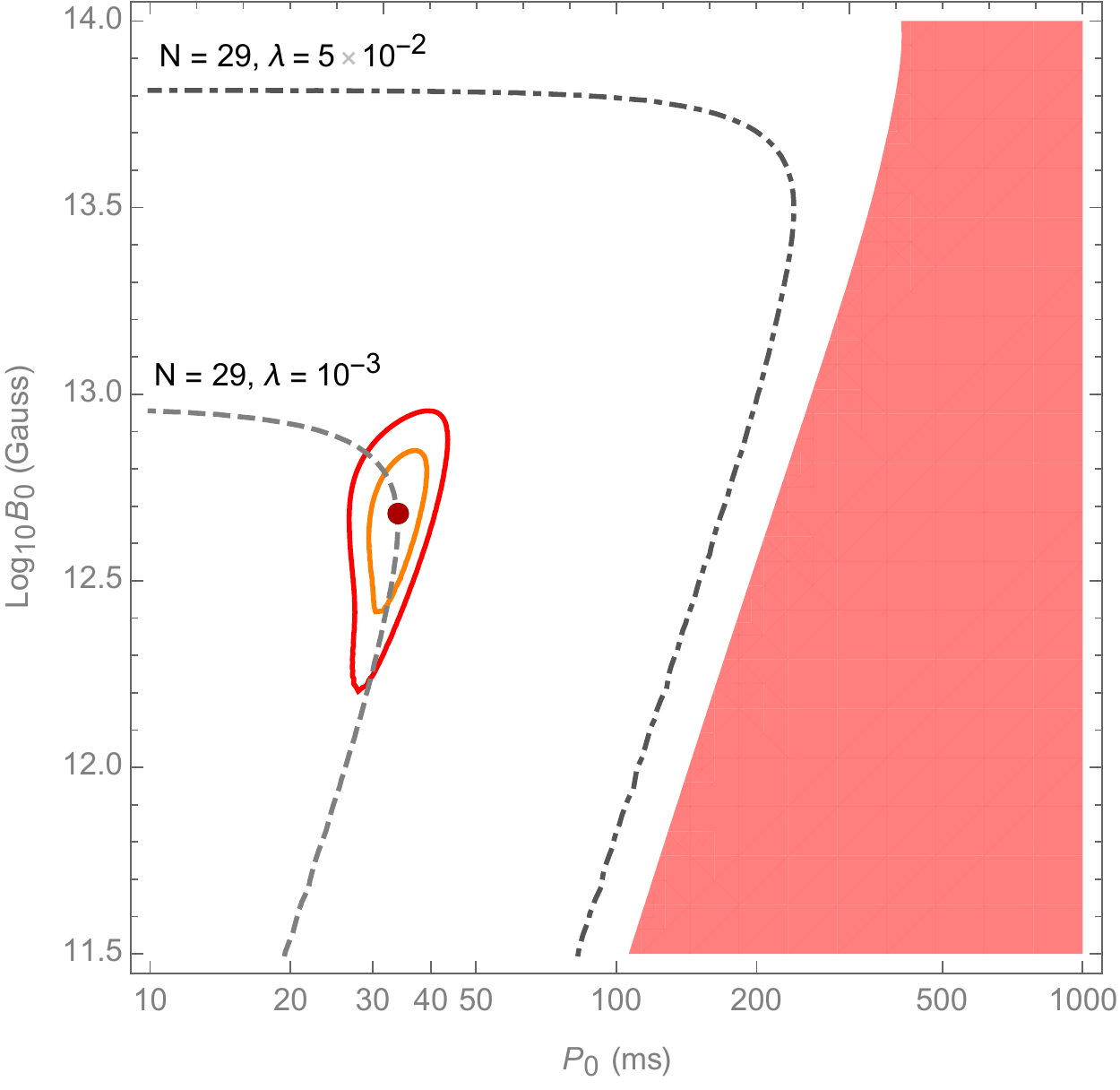}
     \end{subfigure}
\caption{\texttt{Left Panel:}{\em The best fit and the $1\sigma$ and $2\sigma$ allowed regions in the plane $(L_{\rm max},\tau)$. The red shaded area is excluded by the data because corresponds to $N(0.1\Phi_{\rm CRAB})\le 10$ in the assumption of $\lambda = 5 \times 10^{-2}$ which is a large value for the fraction of pulsar spin-down energy converted to TeV emission.} 
\texttt{Right Panel:}
{\em  
The best fit and the $1\sigma$ and $2\sigma$ allowed regions in the plane $(P_0, B_0)$, calculated in the assumption that  $\lambda = 10^{-3}$.
The red shaded area corresponds to $N(0.1\Phi_{\rm CRAB})\le 10$ in the assumption of $\lambda = 5\times 10^{-2}$.}
}
\label{figlr}
\end{figure*}

\subsection{Interpretation in terms of a fading source population}
\label{subsec:fadingsources}

If we consider a fading source population connected with the explosion of core-collapse SN, we can convert the limits on the normalization parameter ${\mathcal N}$ of the source luminosity function into a determination of the fading time-scale $\tau$ through the relationship ${\mathcal N} = R\,\tau (\alpha-1)$. By assuming that the source formation rate $R$ is approximately equal to the SN rate $R_{\rm SN} = 0.019\,{\rm y}^{^-1}$, we get:
\begin{equation}
\tau =  1.8^{+1.5}_{-0.6}\times 10^3\,{\rm y}
\end{equation}
for our reference case, that corresponds to the orange solid line in the left panel of Fig. \ref{figlr}.
Similar values are obtained in the other cases, as reported in Tab. \ref{tabres}.

In the assumption that the observed objects are
PWNe and/or TeV halos which are powered by the formation and the subsequent spin-down of a pulsar, the above value can be used to determine through Eqs.(\ref{P0B0}) the initial period $P_0$ and magnetic field $B_0$ of the considered population. We get the constraints:
\begin{eqnarray}
\nonumber
P_0 &=& 33.5^{+5.4}_{-4.3} \,{\rm ms} \times
\left(\frac{\lambda}{10^{-3}}\right)^{1/2}
\\
B_0 &=& 4.3
\left(1 \pm 0.45\right)\, 10^{12}\,{\rm G} \times  \left(\frac{\lambda}{10^{-3}}\right)^{1/2}
\label{resultsP0andB0}
\end{eqnarray}
that corresponds to the orange solid line in the right panel of Fig.\ref{figlr}.
The small uncertainty for the period $P_{0}$ is connected with the fact that this quantity is determined by the product $L_{\rm max}\tau$ which is relatively well determined by observational data, being the possible variations of $L_{\rm max}$ and $\tau$ anti-correlated.  

We note that inferred magnetic field agrees with the value $\log_{10}(B_0/1G)\simeq 12.65$ obtained by pulsar population studies \cite{FaucherGiguere:2005ny}. 
The inferred period is consistent with the value $P_0 \sim 50\,{\rm ms}$ obtained in \cite{Watters:2010jb} by studying $\gamma-$ray pulsar population. 
The value $P_0\sim 300\,{\rm ms}$ that is obtained from pulsar radio observation \cite{FaucherGiguere:2005ny} is instead excluded by our analysis, unless one assumes that a very large fraction $\lambda \sim 10^{-1}$ of the spin-down power is converted to TeV $\gamma-$ray emission.

The above results are obtained in the assumption that all the sources in the HGPS catalogue with flux $\Phi \ge 0.1\Phi_{\rm CRAB}$  (except those firmly identified as SNRs) are powered by pulsar activity.
A  conservative upper bound for the period $P_0$ can be obtained by considering that no less than 10 of these sources have to be necessarily included in this population, being firmly identified as PWNe or Composite Sources.
The lines $N(0.1\Phi_{\rm CRAB})={\rm const}$ corresponding to a fixed number of sources above the adopted flux threshold $0.1\Phi_{\rm CRAB}$ 
are shown by the gray dashed lines in the planes $(L_{\rm max},\,\tau)$ and $(P_0,\,B_0)$ in Fig. \ref{figlr}.
It can be shown analitically (see Sect.\ref{sec:appendix2}) that $N(\Phi)$ scales as:
$$N(\Phi)\propto \tau \, L_{\rm max}^{3/2} \propto B_{0}\,P_{0}^{-4}\lambda^{3/2}$$
for the limiting case $L_{\rm max}\to 0$, while it scales as:
$$N(\Phi)\propto \tau\, L_{\rm max}^{\alpha -1} \propto B_{0}^{2\alpha-4}\,P_{0}^{6-4\alpha}\lambda^{\alpha -1}$$ 
for $L_{\rm max}\to \infty$.
%
If $1<\alpha<2$, the condition $N(\Phi) = {\rm const}$ always individuates a maximum allowed period $P_0$
(at the transition between the above regimes) whose specific value depends on the fraction $\lambda$ of the pulsar spin-down energy that is converted to TeV $\gamma-$ray emission.
In particular, the red shaded area in Fig.\ref{figlr} can be excluded because it corresponds to $N(0.1\,\Phi_{\rm CRAB})\le 10$ and to the relatively large value $\lambda=5\times 10^{-2}$.
This allows us to obtain the bound $P_0\le500$~ms that can be strengthened if an upper limit for the magnetic field $B_0\le 10^{14}\,{\rm G}$ is introduced.  

\begin{figure}[t]
\includegraphics[width=0.5\textwidth]{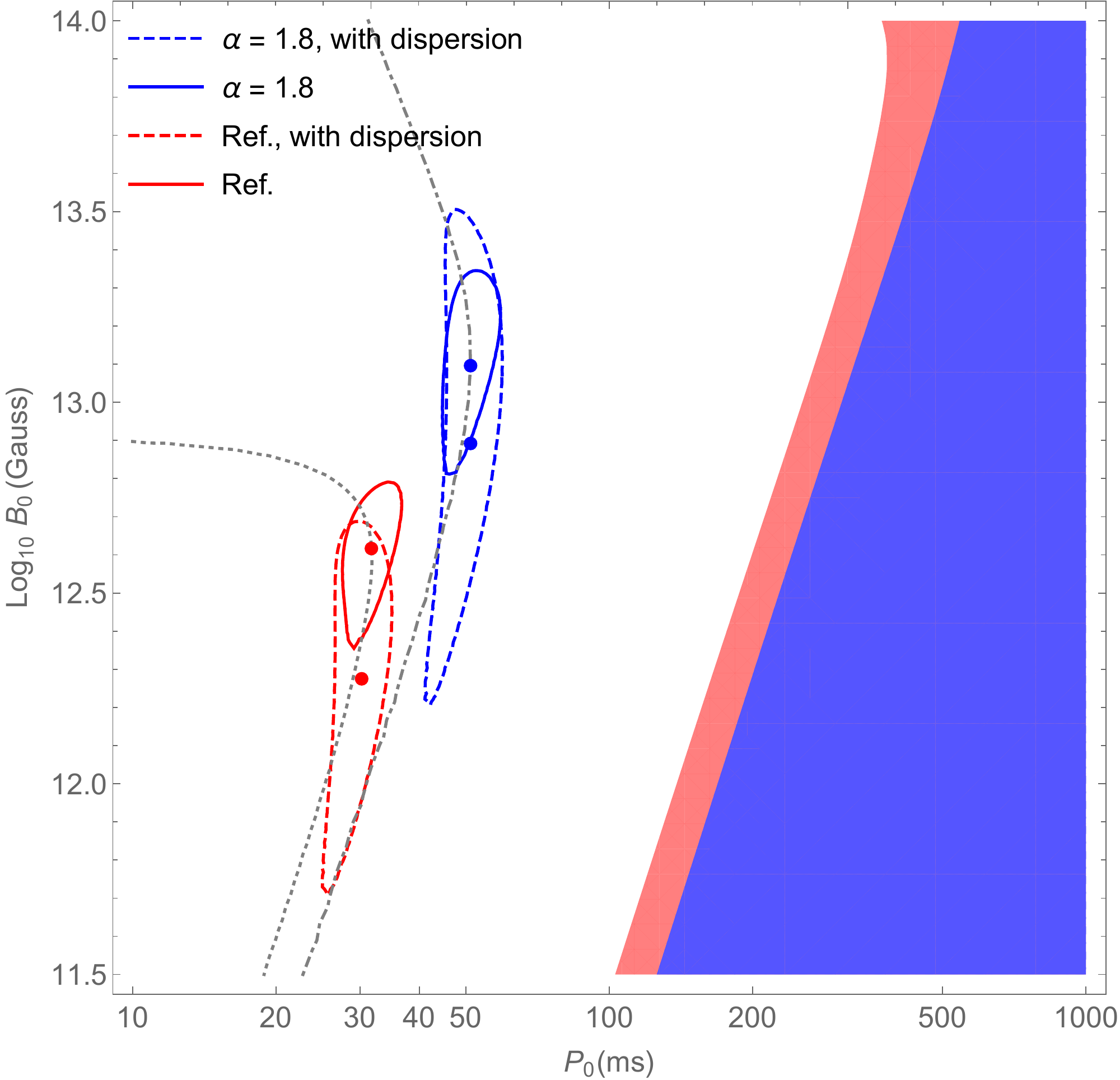}
\vspace{-0.5cm}
\caption{\em 
The best fit and the $1$ and 2$\sigma$ allowed regions regions in the plane $(P_0, B_0)$, calculated in the assumption that the fraction of pulsar spin-down energy converted to TeV emission is $\lambda = 10^{-3}$.
The shaded regions correspond $N(0.1\Phi_{\rm CRAB})\le 10$ in the assumption $\lambda = 5\times 10^{-2}$.
}
\label{dispersions}
\end{figure}

In order to test stability of the constraints given in Eq.~(\ref{resultsP0andB0}), we repeat our calculation for the case $\alpha = 1.8$ obtained by assuming that $\lambda$ is correlated with the spin-down power as suggested by \cite{Abdalla:2017vci}.
In this case, the fading time scale is $\tau= 0.5^{+0.4}_{-0.2}\times 10^{3}$ y, while the initial period and magnetic field are given by:
\begin{eqnarray}
\nonumber
P_0 &=& 51.0^{+8.1}_{-6.4} \,{\rm ms} \times
\left(\frac{\lambda_0}{10^{-3}}\right)^{1/2}
\\
B_0 &=& 12.7^{+9.6}_{-5.8}\, 10^{12}\,{\rm G} \times  \left(\frac{\lambda_0}{10^{-3}}\right)^{1/2}
\label{resultsP0andB0}
\end{eqnarray}
The above results are shown by the blue solid line in Fig. \ref{dispersions} where they are compared with those obtained in the reference case ($\alpha=1.5$).
As a final test, we hypothesize that the initial pulsar periods and magnetic fields are not univocally determined but have log-normal dispersions around preferred values $\widetilde{P}_0$ and $\widetilde{B}_0$ with widths $\sigma_{\log{P}} = \log_{10}(f_p)$ and $\sigma_{\log{B}} = \log_{10}(f_b)$.  
The constraints on $\widetilde{P}_0$ and $\widetilde{B}_0$ that are obtained by choosing $f_{p} = \sqrt{2}$ and $f_{b}= 2$ are displayed by the dashed red and blue lines in Fig.\ref{dispersions}.
We see that the inferred value for $\widetilde{P}_0$ is basically insensitive to assumed dispersions while the preferred magnetic field $\widetilde{B}_0$ is slightly reduced with respect to the reference case, as a consequence of the high-luminosity tail of the source luminosity function that is obtained by assuming $f_p\neq 0$ and $f_b\neq 0$. 

Summarizing, the results displayed in Fig.\ref{figlr} show that the bounds on the initial period and magnetic field do not critically depend on the adopted assumptions, being $P_0$ constrained to the narrow range $25 - 60\,{\rm ms}$ for $\lambda = 10^{-3}$.
The fact that the inferred values for $B_0$ and $P_0$ are consistent with expectations justifies the working assumption that a large fraction of bright sources observed by H.E.S.S. belongs to a population of young pulsars, and supports the hypothesis, formulated e.g. by \cite{Linden} and \cite{Sudoh:2019lav}, that PWNe and/or TeV halos could produce the majority of TeV bright sources in the Sky. 
On the contrary, the large values for the initial period  $P_0\sim 300\,{\rm ms}$ can explain the HGPS results, only if we assume that a limited fraction of observed sources belong to the considered population and/or a consistent fraction of the spin-down energy is converted into TeV $\gamma$-ray emission.  

As a further check of this point, we calculate the expected number of sources in the H.E.S.S. FoV by using the $P_0$ and $B_0$ distributions obtained by \cite{FaucherGiguere:2005ny} from pulsar radio observations, i.e. a gaussian centered in $P_{0}=300\,{\rm ms}$ with standard deviation $\sigma_{\rm P} = 150\ {\rm ms} $, and a log-normal centered in $\log{B_0} = 12.65$ with standard deviation $\sigma_{\log{B}} = 0.55$.
By using the reference value $\lambda = 10^{-3}$, we obtain only $\sim 1$ source above the adopted flux threshold $0.1\Phi_{\rm CRAB}$. In order to reproduce the 10 sources firmly identified as pulsars, we have to assume $\lambda = 1.6 \times 10^{-2}$, while to predict all the 29 sources observed by H.E.S.S. the value of the efficiency $\lambda$ has to be as large as $\sim 5 \times 10^{-2}$.

\section{Conclusions} \label{sec:conclusions}
Recently the H.E.S.S. observatory has completed the first systematic survey of the Galactic plane in the very high-energy domain. Remarkably, the astrophysical nature of the majority of detected sources is still unknown.
In this work, we propose a novel analysis of the flux, longitude and latitude distributions of the brightest sources ($\Phi \ge 10\% \, \Phi_{\rm CRAB}$) of the HGPS catalogue showing that the luminosity distribution of galactic TeV sources can be effectively constrained.

More precisely, by assuming that the luminosity function is described  as a power-law, see Eq.(\ref{LumDist1}), we extract the source maximal luminosity $L_{\rm max} = 4.9^{+3.0}_{-2.1} \times 10^{35}\,{\rm ergs}\,{\rm sec}^{-1}$ and the high-luminosity normalization of the source distribution ${\mathcal N} = 17^{+14}_{-6}$ by fitting HPGS data.
This allows us to determine the total Milky Way luminosity $L_{\rm MW} = 1.7^{+0.5}_{-0.4}\times 10^{37} {\rm ergs}\,{\rm sec}^{-1}$ in the energy range $1-100\,{\rm TeV}$ and the total Galactic flux in the H.E.S.S. FoV given by  $\Phi_{\rm tot} = 3.8^{+1.0}_{-1.0}\times 10^{-10}\, {\rm cm}^{-2}\, {\rm sec}^{-1}$. 
The luminosity $L_{\rm MW}$ is only a factor $\sim 4$ smaller than that obtained in the energy range $1-100\,{\rm GeV}$ by fitting Fermi-LAT 3FGL and 1FHL catalogue.
In addition, the total source flux is relatively large, implying that unresolved source contribution is not negligible (about 60\% of the resolved signal measured by H.E.S.S.) and potentially responsible for a large fraction of the diffuse-large scale gamma-ray signal observed by H.E.S.S. and other experiments in the TeV domain. The unresolved contribution can e.g. explain the excess reported by Milagro at 15 TeV \cite{Atkins:2005wu}.
Moreover, we consider the possibility that the bright sources observed by H.E.S.S., which are not firmly identified as SNRs, are powered by pulsar activity, like e.g. PWNe and TeV halos. We evaluate the constraints on the physical properties of the pulsar population that follow from this hypothesis.
For our reference case, assuming that the fraction of the pulsar spin-down energy converted in TeV photons is $\lambda=10^{-3}$, we obtain the best-fit values $P_0 = 33.5^{+5.4}_{-4.3} \,{\rm ms}$  and $B_0 = 4.32 \left(1 \pm 0.45\right)\, 10^{12}\,{\rm G}$, the initial spin period and magnetic field, respectively.
The above constraints are consistent with the $B_0$ values obtained in \cite{FaucherGiguere:2005ny} and $P_{0}$ constrains described in \cite{Watters:2010jb} by studying the gamma-ray pulsar population.

Finally, by considering that 10 sources in HPGS catalouge have been firmly identified as PWNe and considering $\lambda \le 5\times 10^{-2}$ as an upper bound for efficiency of TeV emission, we obtain that the intial spin-down period of the considered pulsar population is costrained to be $P_0 \le 500 \,{\rm ms}$.

\vspace{1.0cm}

\section*{Acknowledgements} \label{sec:acknowledgements}
The Authors are grateful to Paolo Lipari for fruitful collaboration, critical discussion and useful suggestions for the completion of the manuscript.
The Authors are grateful to Pierre Cristofari for discussions and helpful comments.
This work was partially supported by the research grant number 2017W4HA7S ''NAT-NET:
Neutrino and Astroparticle Theory Network'' under the program PRIN 2017 funded by the Italian Ministero dell'Istruzione, dell'Universita' e della Ricerca (MIUR).

\appendix*
\begin{appendices}

\section{}
\label{sec:appendix}

\subsection{Likelihood definition}
\label{sec:appendix1}
In order to determine the maximal luminosity $L_{\rm max}$ and the normalization $\mathcal N$ of the luminosity function, see Eq.(\ref{LumDist1}), we use the maximum Likelihood technique. 
The H.E.S.S. catalogue contains 78 sources with their Galactic coordinates ($b_{\rm i}$, $l_{\rm i}$), the observed fluxes $\Phi_{\rm i}$ in the energy range $1-100\;{\rm TeV}$ and the respective uncertainty $\delta\Phi_{\rm i}$.
In our work we considered only the $32$ brightest sources with a flux above $1\;{\rm TeV}$ larger than $0.1\Phi_{\rm CRAB}$
for which the H.E.S.S. catalogue can be considered complete.

Given this data set we define an unbinned Likelihood function $\mathcal{L}$, according to:
\begin{equation}
\log{\mathcal{L}} = - \mu_{\rm tot} + \sum_{i} \log{( \mu_{\rm i})}
\end{equation} 
where $\mu_{\rm tot}$ represents the number of expected sources, while $\mu_i$ is the probability to observe an object with coordinates ($b_{\rm i}$, $l_{\rm i}$) and measured flux $\Phi_{\rm i}$. 
These quantities are calculated by considering that the source distribution per unit of flux $\Phi$ and solid angle $d\Omega$ is given by:
\begin{equation} 
\label{mu}
\mu(b, l, \Phi) 
= \int dr \; 4 \pi r^4 \langle E\rangle\ Y(4 \pi r^2\langle E\rangle\ \Phi)\ \rho(r, b, l)
\end{equation}
with the functions $Y(L)$ and $\rho({\bf r})$ defined in Sect.\ref{sec:method}. The parameter $\mu_{\rm tot}$ is obtained by integrating the function $\mu(b, l, \Phi)$ in the HESS FoV and in the flux range $\Phi\ge 0.1 \Phi_{\rm CRAB}$.
The coefficients $\mu_{\rm i}$ are obtained as:
\begin{equation} \label{mueff}
	\mu_i = \int d\Phi \;  \mu(b_{\rm i}, l_{\rm i}, \Phi) P(\Phi_{\rm i}, \Phi, \delta \Phi_{\rm i}). 
\end{equation}
where the function $P(\tilde{\Phi}, \Phi, \sigma)$ represents the probability that 
the {\em measured} flux $\tilde{\Phi}$ is obtained for a source emitting the {\em real} flux $\Phi$. We assume that this can be described by a Gaussian with a dispersion $\sigma$ equal to the uncertainty of the measured flux, i.e. 
\begin{equation}	P(\Phi_{\rm i}, \Phi, \delta \Phi_{\rm i}) = \frac{1}{\sqrt{2 \pi \delta \, \Phi_{\rm i}^2}}\; \exp\left[{- \frac{(\Phi - \Phi_{\rm i})^2}{2\, \delta \Phi_{\rm i}^2}}\right].
\end{equation}
Finally, the best fit values and the allowed regions for the parameters in our analysis are obtained by studying the $\chi^2$ behaviour, defined according to:
\begin{equation}
\chi^2= -2 \log{\mathcal L}.
\end{equation}

\subsection{The flux distribution}
\label{sec:appendix2}

The flux distribution can be calculated as:
\begin{equation}
\frac{dN}{d\Phi} = \int dr\; 
4\pi r^4
\langle E \rangle \;
Y(4\pi r^2 \langle E \rangle \Phi)\;
\overline{\rho} (r) \;
\label{dNdphi}
\end{equation}
where $\overline{\rho}(r) \equiv \int_{\rm FoV} d\Omega \; \rho(r, {\bf n})$ is the source spatial distribution integrated over the longitude and latitude intervals probed by H.E.S.S.
Note that integration in Eq.\eqref{dNdphi} is limited to the distance range $ r \le D(L_{\rm max},\Phi)$ where $D(L,\Phi) \equiv \sqrt{L/ 4\pi  \langle E \rangle \Phi}$ represents the distance below which a source with intrinsic luminosity $L$ produces a flux larger than $\Phi$.
Moreover, in the assumption that sources have a physical dimension $d$, one also has a lower integration limit $r \ge d/\theta_{\rm max}$, where $\theta_{\rm max}$ is the maximal angular dimension that can be probed by H.E.S.S.
%


The function $dN/d\Phi$ can be calculated analytically in the two limit cases $L_{\rm max}\to \infty$ and $L_{\rm max}\to 0$.
For $L_{\rm max}\to \infty$, hence $D(L_{\rm max},\Phi) \to \infty$, the function is:
\begin{equation}
\begin{split}
\frac{dN}{d\Phi} =& \; R \; \tau \; (\alpha-1) \; L_{\rm max}^{\alpha-1} \;\Phi^{-\alpha}\\
&\times \int_{0}^{\infty} dr\; 
(4 \pi \langle E \rangle)^{1-\alpha} \; r^{4-2\alpha} \;
\overline{\rho} (r) ;
\end{split}
\label{dNdphiInf}
\end{equation}
here the integral is only dependent on the coordinate $r$ and is therefore a constant. The dependence on $\Phi$ is only given by the term $\Phi^{-\alpha}$; the total number of sources $N(\Phi)$ above a flux $\Phi$, which is shown in Fig.\ref{fig2}, is therefore proportional to $\Phi^{-\alpha+1}$.
For the limit case $L_{\rm max}\to 0$, hence $D(L_{\rm max},\Phi)\to 0$, the integral over $r$ is extended to a small region where the distribution function $\overline{\rho}(r)$ can be considered constant and equal to its value at $r=0$, i.e. $\overline{\rho}(r)\simeq \overline{\rho} (0)$. We thus obtain:
\begin{equation}
\begin{split}
\frac{dN}{d\Phi} \simeq& \; (4 \pi \langle E \rangle)^{1-\alpha} \; \overline{\rho} (0)\; R \; \tau \; (\alpha-1) \; L_{\rm max}^{\alpha-1} \;\Phi^{-\alpha} \\ 
&\times \int_{0}^{D(L_{\rm max},\Phi)} dr\; r^{4-2\alpha} \\
=& \; \overline{\rho} (0) \; R \; \tau  \left(\frac{\alpha-1}{5-2\alpha}\right) \left(\frac{L_{\rm max}}{4 \pi \langle E \rangle}\right)^{\frac{3}{2}} \;\Phi^{-\frac{5}{2}} ;
\end{split}
\label{dNdphiZero}
\end{equation}
The cumulative function $N(\Phi)$ is therefore independent from the index $\alpha$ considered and is proportional to $\Phi^{-\frac{3}{2}}$ as it is shown in Fig.\ref{fig2}.

\end{appendices}

\end{document}